\documentclass[12pt]{article}
\makeatletter

\@addtoreset{equation}{section}
\makeatother
\usepackage{amsfonts}
\usepackage{eucal}

\begin{document}

\title{An Introduction to time generation on  Algebraic Quantum Field Theory}
\author{Tadashi FUJIMOTO\thanks{Department of Philosophy, Ryukoku University 600-8268 Kyoto, Japan. e-mail: fujimoto-t@let.ryukoku.ac.jp}}

\date{January 10, 2020}
\maketitle

\begin{abstract}
In the early 2000s, the study of time operators advanced as one of the methods to understand the problem of time as mathematical science. However, the starting point for the time operator is to understand time as a problem of observation (the“survival probability”of particles), and today, even after the issue of representation on time operator has concluded, the question of philosophical interpretation still exists. Furthermore, when it comes to the question of how time ``generation (emergence) '', the method of time operators has its limitations. Regarding the generation of time, `` symmetry breaking '' in particle physics seems to be closely related.
\end{abstract}


\section{Intriduction}
It is anticipated that time operator will play an important role in physics, especially quantum physics, and is being discussed in many articles and textbooks today. Unlike the age of W. Pauli who once argued time operator negatively, it is sometimes extended to the discussion of non-commutative space-time theory. However, the time operator has complicated appearance, not only in its mathematical representation theory, but also depending on the topology and the domain in which it operates. In addition, this operator was originally revisited as a problem of ``survival probability"of particles (related to the scattering problem) based on uncertainty relations, and directly linked to the discussion of space-time non-commutativity would have to be careful.

The most important problem is that time operator is linked to Hamiltonian of the system. The Hamiltonian, as is well known, controls the translation of time by the Noether's theorem. When a``time"range is fixed, the Hamiltonian determines the symmetry of the system with respect to time through showing energy as the conserved quantity of the physical system. Therefore, a particular time operator can only be applied to the behavior of particles and fields in the local world of one of ``reference systems''\footnote{For example, macro measurable physical systems such as local relativistic inertial systems.}. In other words, the theory of time operator is not generally applicable to non-inertial systems where the Hamiltonian form cannot be used. In addition, the theory of time operators is, by itself, somewhat narrow in terms of how time``comes''.

In this paper, taking consideration of the above problems, I present another view on the generation of time and attempt to combine it with the discussion of time operators. At that time, I would like to deal with two specific issues (entanglement and cluster decomposition,) related to the quantum field theory.

\section{Ontology and Epistemology }
First, let's start by discussing the uncertainty relations and its interpretation.

As is well known, in the recent development of the study on uncertainty relations (the process of finding Ozawa's inequality), as M.Ozawa stated, Heisenberg's discussion includes not only the aspect of so-called measurement accuracy but also the aspect (Kennard's inequality) that emphasized the non-commutativity of operators from the mathematical meaning of the standard deviation [16]. What Ozawa pointed out at this time is the difference between the objects in physics where measurements are taken and the object analyzed in the context of mathematics. Ozawa thinks that in the process of Heisenberg's uncertainty relations being refined as mathematics the reflection on observations and measurements in physics was not taken into account so much and that the relations has been buried in operator algebras' problems. In recent years, the context of physics has been reviewed through the detection of gravitational waves. This is an epistemological approach to uncertainty relations. In other words, when we regard the problem of non-commutativity on operators which is a kind of mathematical approach as an ontological approach, this mathematical approach is overlooked in the problem of physical observation. An epistemological approach is to correctly evaluate its physical context dependency.

Indeed, in the context of mathematics (ontology), it does not matter where the non-commutative state of the operator signifies the phenomenon or observation process.
The strength of mathematical theory lies in its generality that it is not affected by the physical context. However, when mathematics comes to the application of the physical theory, there is not intrinsic, a priori path to physics phenomena in mathematics itself. In order to discuss this precisely, it is necessary to deal with a rather large theme that takes into account the disciplines and properties of mathematics and physics. Therefore, it cannot be mentioned in this paper. However, physical mathematics as a``tool'' for physics and the mathematics as``mathematics"cannot be spoken at the same level. It is worth pointing out here that the history of mathematics and physics lies underneath [8, 20].

Mathematical approaches (ontological approaches) are not mistaken as approaches in uncertainty relations. Exact proof has been given to the extent that it is discussed as mathematics. What is important is that the elements themselves that causes the uncertainty is also caused by various errors and problems of detection accuracy in the observation and detection processes of the physical system. Such concrete elements are not elements that appear in a priori from the beginning in mathematics. Regarding quantum entanglement, a similar relationship exists between the well-established theory of mathematics and the experimental data that has recently become demonstrable with the progress of optical (engineering) technology.

Considering the entanglement problem as mathematics, for example, as an operator algebraic approach, as follows.

\medskip

Let's $A$, $B$ be $C^{*}$ algebras, and,  $\omega^{A}:=\omega(a\otimes{1})$,        $\omega^{B}:=\omega(1{\otimes}{b})$, then the entangle state is in a broad sense the state which is not be described as like $\omega(a\otimes{b})=\omega^{A}(a)\bullet\omega^{B}(b)$. (the state which is not be represented as a product state) [5].

\medskip

However, exact quantum entanglement\footnote {Sometimes called ``quantum correlation'' or ``quantum entanglement''} is stronger than the classical correlation and the measurement result does not depend on how to take the basis on Hilbert space. The mathematical proof that certain measurements of correlated spatially separated systems are necessarily nonlocal is given by J.Bell and is called Bell's inequality. The entanglement state is not discussed so strictly in this paper. However, while the entanglement state is discussed as a mysterious problem in philosophy, it is overwhelmingly more expensive than the non-entanglement state ($ 2 ^ {n} / 2n$: ratio). Note that entanglement state is not a peculiar state.

In mathematics, the separation of quantum entanglement is done by means of``singular value decomposition (Schmidt decomposition) ''. This is an application of a kind of dilation theory, for example by adding degrees of freedom to matrix operations [14, 18].  Neither this entanglement nor its decomposition is paradox at all. However, in quantum physics issues, entanglement is sometimes taken up as a so-called mystery problem because the``causality''problem comes to the forefront in areas that are spatially separated.

However, from the mathematics (and information theory) point of view, the relativistic requirements have not been violated at all; rather, entanglement appears to be a paradox, not from both ends receiving information, but from a third perspective (from transcendental viewpoint), which is the cause of what is understood as``non-locality" \footnote {To put it a little further: correlation and causality are different, but they are often confused If we notice this fact, this issue is not a paradox.}.

In the following, I will discuss the issue of ``time and causality'' in relation to quantum entanglement, which will provide a starting point for understanding the generation of time.

\section{Time and Causality}

Mathematics requires a chain of assumptions and consequences in the proof process ( causality in a broad sense). But``time" causality does not come to the forefront of mathematical proofs. The consequences of natural sciences, such as physics, are usually temporal consequences because natural phenomena are correlated with time parameters. In fact, there is no conclusion in today's philosophy of science whether time is based on causality or whether causality is based on time. In this regard, when one considers causality in terms of the ontological and epistemological relations discussed above, for example, the mathematical approach may be roughly approximated by C.G.Hempel's ``explanatory theory." However, because mathematics is not exactly the same as natural science, there are many points that are difficult to interpret in explanatory theory [11].

 I would like to emphasize the mathematical point of view (ontological point of view) here, so that the causal problem of time does not appear. There is only a correlation in non-temporal sense. Mathematically, states can not only be represented by tensor products, but they can always go to product states by applying dilation theory. The entanglement problem appears to be a paradox because the epistemological view of the phenomenon is understood in the everyday concept of time. In other words, it is as if the demand for special relativity is violated and we fall into the illusion that information (a kind of energy) can be transmitted instantaneously. As we shall see later, this is due to the extended interpretation of one local reference system. Therefore, the entanglement problem is completely compatible with Algebraic Quantum Field Theory (AQFT), i.e, Local Quantum Field Theory.

Next, in order to delve into these problems, let's interpret from the viewpoint of the cluster decomposition theorem, which is also a entanglement problem in the broad sense.

\medskip

As a relationship between mathematics and physics, various cluster decomposition theorems serve as indicators, and problems on time are hidden there.

Let's take an example. This is the Cluster theorem for the vacuum of local fields in special relativity. This is called the Jost-Lehmann-Dyson theorem (1962). An improved version was released by K.Fredenhagen in 1985. If there is a spectral gap between the ground state $\Omega $ and the excited state (where the operator is acting), the following inequality holds \footnote {For details omitting other conditions, see the original paper [9]}.

$|(\Omega, AB\Omega)-(\Omega, A\Omega)(\Omega,B\Omega)|{\le}{e^{-m\tau}}(\|{A^{*}}\Omega\|\|{B}\Omega\|+\|{A\Omega}\|\|{B^{*}\Omega}\|)$ (Assuming translation invariance).

When evaluating the correlation term using the cluster term (right side), (if the physical quantity is spatial) the spectral gap $m$ which is energy information, and the positive time parameter are constant terms. It can be said that the correlation term on the left side measures the distance of the local area at a fixed time. Time measures distance (for example, Yukawa potential). That is, the left side is a broad entangled state, which is evaluated by a non-entangled state (cluster state). At this time, time appears in the evaluation term as a problem of observation (as an epistemological problem). Then, the origin of time $\tau$ here will be an issue. There is an important perspective when considering this issue.   

\medskip

Regarding $e^ {-itH}$ and heat kernel $e^ {-\beta {H}}$, $ (\beta>0)$, which generally have a semi-group structure and control spatial translation, A.Arai states that the time evolution and thermal equilibrium state are regarded as a segment of one object (in this case, a semi-group and a group structure) [1]. This point is extremely important. In the evaluation formula using the cluster is the same as $e^ {-\beta {H}} $, $(\beta>0) $ is the same as if only the appearance is considered. 

In the case of AQFT, for a physical quantity $ \mathcal {A} $ ($D$) in a limited spatial and  a limited time (finite space-time area $D$), the axioms of local physical quantities are assumed to be (domain) monotone property, (Poincare) covariance, (spatial) locality, and (algebraic) generating property [4]. In this case, the bounded space-time is assumed to be a macro (classical) space-time based on special relativity. If we consider here that the von Neumann algebra generated by a physical quantity localized in this bounded region is of type I\hspace {-.1em}I\hspace {-.1em}I, then it is a misconception that quantum entanglement breaks the locality of AQFT. Bounded areas cannot communicate beyond relativistic demands. However, the transmission of one is deterministic (instantly) deterministic of the other. There is a kind of misunderstanding about understanding time and causality. It is important to clarify this point.

\section{Aspects of micro-macro time in quantum physics}

Here, I would like to set``the aspect of time in quantum theory''as follows.

\medskip

・(A) External time: Macro time measured during an experiment.

・(B) Observation time: Time related to the observation itself (relationship with the observer)

・(C) Internal time: Time inherent in the micro object itself.

\medskip
This classification follows W.Hagenberg's Copenhagen interpretation. (A), (B), and (C) are read as follows [10, 21], furthermore, add ontological and epistemological viewpoints in AQFT [15]

\medskip

・(A) classical physical time (phenomenoning time).

・von Neumann algebra as connection of (A)-(B) $ \mathcal {A}$ ($D$)

・(B) Time for observation (Instrument, Arverson spectrum)

・(Thermodynamic) phases (pure phase, mixed phase, superselection rule) due to the sector of von Neumann algebra as connection of (B)-(C)

・(C) non-phenomenal time (quantum time).

\medskip

By the way, in (B), the quantum object is correlated with the macro observation detector [15, 17]. However, the interference term of the observation process (first-class process i.e, von Neumann-L\"{u}ders type) by the density operator can be eliminated by increasing the degree of freedom of the observation detector. An instrument is defined as a complete positive mapping of trace preservation. In this case we use the Positive Operator Valued Measure (POVM). The instrument measurement description holds the correlation between the observation device and the target micro-system (with a certain degree of freedom).

Details of the Arverson spectrum can be found in the text by M. Takesaki [22]. In short, it is a spectrum whose operator algebra representation extends beyond the commutative $C^{*}$ to a commutative Banach algebra, where the spectrum is defined by the zeros of the Fourier transform. This extends spectral theory from normal and unitary operators in Hilbert space to a slightly wider class of operators. After setting up the stage equipment as described above, I would like to do a little philosophical rearrangement on the positioning of space and time.
 
Heisenberg's composition is in the genealogy of Plato, a position that places time and space between the non-sensitive world (Idea world) and the phenomenal world (event world). In fact, Heisenberg himself acknowledged the influence of Plato.

We have discussed the heat kernel involved in the Hamiltonian above. Speaking of the issue of time and heat, from the point of view of philosophy, for example, consider the following questions: ``Does consciousness work in a completely still world ?'', ``Is there time where consciousness doesn't work ?" Now what ?

\section{Background of time generation-symmetry breaking}

Here is an example that provide important clue when considering the generation of time. The hint is the integral decomposition and symmetry breaking of BEC (Bose-Einstein condensation) state [1]\footnote{Weyl operator $\{ W(f)| f\in{\oplus}^{N_{i}}L^{2}(R^d)\}$ and the algebra generated by it are assumed}.
In general, the BEC state $\omega_ {BEC} (W (f))$ does not have spatio-temporal cluster, but spontaneous symmetry breaking  realizes clustering ($\omega_ {r, \theta} $:
perform change  of variables).

Via Araki - Woods representation, we create GNS construction\footnote {[4]. Gelfand-Naimark-Segal construction.}, ($ \omega_ {BEC} (W (f)) = (\Omega_ { BEC}, W_ {BEC} (f) \Omega_ {BEC} $), and assuming the non-equivalence of the parameters after change of variables, all cyclic representations are non-equivalent. Thus, the representation of direct product on fibers is as follows.

$\textit{H}_{BEC}=\int^{\oplus}_{[0,\infty)\times[0, 2\pi]}\textit{H}_{B}d\mu(r, \theta)$,

$W_{BEC}(f)=\int^{\oplus}_{[0,\infty)\times[0, 2\pi]}W^{r, \theta}(f)d\mu(r, \theta)$

\medskip

Spontaneous symmetry breaking occurs in the case that the physical constants of time evolution are invariant. Thus, overall gauge symmetry with respect to vacuum is preserved, but even so, for each vacuum, the GNS representations are non-equivalent to each other, and clustering holds. This situation is related to the KMS state\footnote{[22]. Kubo-Martin-Schwiner condition.} with different temperatures $\beta $ based on the theory of the type I\hspace {-.1em}I\hspace{-.1em}I factor. The time parameter $ t $ can be interpreted to occur when some sort of breaking concept overlaps this ``fiber". This leads to the interpretation that the sequence of fibers in the mathematical (ontological) sense gives rise to the perception of time.

Let us consider the generalization of the above, that is, the case of explicit symmetry breaking, in which the physical constants that determine the time evolution of the system move on a certain fiber. Flavor quantum number breaking, mass difference between two kinds of $\pi$ mesons in electroweak theory, and neutrino oscillation lead to this breaking. As we can imagine from CPT breaking, we can assume a substratum structure that breaks the continuity of time, violating the standard model and breaking the energy conservation derived from the invariance of time translation (with respect to the Hamiltonian).

A time evolution of a physical system must be different for each type of fiber (although it is not clear whether the expression``fiber"is appropriate for explicit symmetry breaking) [15]. If so, further below spontaneous breaking, there may be some mechanism that governs continuous infinity and changes to physical constants, such as generating  time of a certain system. To approach it with AQFT, it is necessary to base the bounded region from the observation side (further development of scale conversion; scaling algebra). 
In other words, it is necessary to consider not the real parameters of the local system in special relativity but the connection with the structure of the base layer. Considering the problem of quantum entanglement again, the following view may be possible.

In a communication, interpreting that one side receives the information can be interpreted as a symmetry breaking  (a kind of clustering is established). It is not a paradox due to temporal causation, but a correlation is determined. It is the generation of a reference system, which determines the time evolution of the local system, or the thermal structure. The time parameter in the time representation and the layer of time (as like fiber) generated each time have different ranks (or phases). However, what we can usually observe is the time of the reference system and its complementary thermal phenomena. The error of globalizing the local time or the error caused by fixing the parameters of thermal phenomena is the entanglement paradox. Behind the thinking of entanglement as a causal break, there may be a process of generating time that satisfies local system stability.

\section{Time operator and its implication}

Here, we would like to review the time operator in order to reconsider the problem of time.
Historically, observations of energy in the unsteady state have shown that there is a dispersion relationship between time and energy. Heisenberg presents $\delta$ {E} $\delta $ {T} $ \sim {h} $, ie the time-energy uncertainty relation, through the Stern-Gerlach experiment.

$\Delta{T}\Delta{E}{\ge}\frac{1}{2}\hbar$,  $\Delta$ is the standard deviation [12].

Time uncertainty can be understood as an epistemological issue because it is physically a matter of measurement [6]. However, in recent non-commutative geometry of space-time and AQFT, when considering the system of operator algebras on curved space-time, mathematics requires ontological arguments through the operators themselves. I stated at the beginning of this article that I was somewhat skeptical about whether this view were right.
By the way, the first time operator constructed historically is by Aharonov and Bohm, which constitute one-dimensional time operator, called the Aharonov-Bohm time operator and written in the form: $T = \frac {1} {2m} (Q {P^ {-1}} + {P^ {-1}} Q) $.
$ Q $ is a position operator, and $ P $ is a momentum operator.

This operator has a canonical commutation relation (CCR) with the free particle Hamiltonian $ H = \frac {P ^ 2} {2m} $. That is, $ [T, H] = i $. ($\hbar=1$) [3]. The meaning of `` $ T $ '' as a (generally) symmetric operator has been solved today in terms of representational problems [2].

\medskip
The next most important issue of a time operator is its relation to a time parameter in a classical physics sense. Since the time operators are generally symmetric operators, they are known from operator theory that their spectra are in a set of complex numbers. In this paper, I would like to go one step further and think from the point of view of the instrument. Unitary time evolution with $ e^ {-itH} $ is used for the instrument itself. However, it can be viewed as describing a single temporal structure of motion after generation (in a reference system). The question is how to get that time back into the area where time is generated.

One view is based on the idea of ​​C. Rovelli's Partial Observables[19], which is based on the idea of ​​a positive operator measure using dilation theorem. As a theory of operator algebras, a method using the weight theory is conceivable [7]. In other words, considering the size of the Hilbert space, the characteristics of the spectrum established in a large Hilbert space (for example, a real number spectrum by a multiplication operator) is lost in a small size Hilbert space (spectrum becomes a set of complex numbers). The POVM used in the instrument is used there\footnote {The following is an example of POVM.

For any $t\in [0, 2 \pi] $, define POVM $F(S) $ as follows:
$$F(S) = \Sigma_ {n, m\ge0} \frac {1} {2\pi} {\int} e^ {i (nm) t}dt|n\rangle {\langle} m | $$
  (mod $ 2\pi $)

$$ T = \int ^ {2\pi} _ {0} t\lambda {dF} (\lambda) = \Sigma_ {m\neq {n} \ge0} \frac {1} {i (nm)} | n\rangle {\langle} m | + \pi {I} $$

This is viewed as a generalized spectral measure.}.

But there's a big problem here too.
Unlike other operators and parameters, describing the evolution of a time operator with a time parameter (for example, a semi-group notation such as $ \alpha_ {t} $) is a logical cycle. POVM also uses parameters as a time shift, and cannot describe the generation of time. This is the same problem in the case of the instrument. However, because ``Weight" is more general than GNS using a positive linear functional, it may open the way to non-commutative stochastic time. Thereby in addition, we may find a new relationship  to a spectrum of von Neumann algebras of type  I\hspace{-.1em}I\hspace {-.1em}I. Transformation from heat to time, or time basis from physical variables that do not go through time, is needed. If so, we can also cast doubt on the thesis that time is a priori.

\medskip

For the purposes of this paper, at least time as the parameter $ t $ is not considered to be the base of time as I discussed in my book,

1) Historically, this parameter was invented and discovered shortly before Galileo's era (as of N. Oreme) as a convenience of spatial movement.

2) The (non-commutative) problem of time cannot be ignored when considering the gravity problem.

3) Mathematics   is not linked to the concept of time directly.

4) Space-time symmetry (external symmetry) and internal symmetry (related to quantum numbers, etc.) are essentially different.

\medskip

Therefore, I would like to finally introduce a new interpretation of time based on future developments.

\section{Outlook}

The new perspective is the deep meaning of $ e ^ {-\beta {H}} $ and $ \Delta ^ {it} $ for the modular operator $ \Delta $ in Tomita-Takesaki theory. Or, in $ {\beta} =-1 $, it means $ \Delta^ {it} = e^ {{it} ln \Delta} $$ \to $$ e^ {itH} $ (The automorphism group is limited to the modular automorphism group).

The guiding philosophy is``symmetry breaking ''and``generation of time'' which are often discussed in this paper. Specifically, there is a relationship between CPT and modular conjugate operator. Here, it is necessary to consider the connection with the modular Hamiltonian $ ln\Delta $ (the Lorentz linear transformation matrix S). Alternatively, considering the decomposition of entanglement (cluster decomposition) first, it cannot be denied that increasing the dimension of the target system may lead to the essence of time
\footnote {Superstring theory with supersymmetry would be a useful working hypothesis.}.
The entangled state moves to the product state by increasing the dimension of the space. If the entangled state corresponds to the symmetry, the product state can be regarded as a state of broken symmetry or a state without symmetry. Based on sector theory, entangled states could be defined as spatially limited states of product states. If such an assumption is possible, it may be inferred that the archetypal time created from the symmetry-breaking state is expressed as a time parameter in the non-local correlation breaking (cluster state) in a limited dimension. To accurately describe such inferences, we will have to investigate the relationship between analysis of the extra dimensions of operators and symmetry breaking. I will discuss these issues in other papers.

When we consider the generation of time and the continuity of time, or the relationship between time and causation as a philosophy, it will be  important to consider the world view that there is a symmetry breaking first and that the symmetry itself  emerges late\footnote {P. Curie stated that ``asymmetry cannot arose from symmetry" [13]. Symmetry: static (harmonic) versus asymmetry: motion (chaos).}. AQFT can be a powerful method because the problem of time is linked to aspects of generating time.
 
How is mathematics about time possible? Where is time? Where does``time"or something as like``fiber"become time? 

It is difficult to find a kind of hierarchical and vertical structure for the description of classical physics, and that role will have to be played by philosophical idea and mathematical quantum physics today. There is a need to correlate the existence of mathematics with the sensual world and harmonize it with the epistemological view of physics.

\section*{Acknowledgments}

This paper is based on a manuscript presented at the research meeting commemorating the retirement of Prof. Asao Arai held at Hokkaido University in March 2019. 

Here, I would like to add that there is discussions with Prof. Izumi Ojima at Kyoto University and Prof. Kraus Fedenghagen at University of Hamburg, which I have taken care of so far. Special thanks to the teachers.

\end{document}